\documentclass[a4paper, 10pt, conference]{cssconf}
\IEEEoverridecommandlockouts    
\usepackage{epsfig} 
\usepackage{amsmath} 

\title{\LARGE \bf
Converting Biomechanical Models from OpenSim to MuJoCo
}

\author{Aleksi Ikkala, Perttu H\"{a}m\"{a}l\"{a}inen
\thanks{This work has been supported by Academy of Finland grant 299358.}
\thanks{A. Ikkala and P. H\"{a}m\"{a}l\"{a}inen are with the Computer Science Department of Aalto University, Espoo, Finland (correspondence to {\tt\small aleksi.ikkala@aalto.fi}).}
}

\begin{document}
\maketitle
\thispagestyle{empty}
\pagestyle{empty}

\begin{abstract}


OpenSim is a widely used biomechanics simulator with several anatomically accurate human musculo-skeletal models. While OpenSim provides useful tools to analyse human movement, it is not fast enough to be routinely used for emerging research directions, e.g., learning and simulating motor control through deep neural networks and Reinforcement Learning (RL). We propose a framework for converting OpenSim models to MuJoCo, the de facto simulator in machine learning research, which itself lacks accurate musculo-skeletal human models. We show that with a few simple approximations of anatomical details, an OpenSim model can be automatically converted to a MuJoCo version that runs up to 600 times faster. We also demonstrate an approach to computationally optimize MuJoCo model parameters so that forward simulations of both simulators produce similar results.

\end{abstract}

\section{INTRODUCTION}
\label{sec:introduction}

OpenSim (\cite{delp2007, seth2018}) is a physics simulator extensively used by biomechanics researchers. This community of researchers have also created many human and animal musculo-skeletal models, often based on cadaver studies. The unparalleled anatomical accuracy of various models have been validated in several papers (for instance, \cite{saul2015, arnold2010}).

Although OpenSim models excel in accuracy, the simulator lacks in speed: a forward simulation of a movement that lasts a few seconds can take minutes to run on a complex model comprised of tens of muscles and joints. Therefore, running OpenSim models on another physics simulator might provide enough speed-up to use these anatomically accurate models for e.g. machine learning or animation research. 

MuJoCo is a fast and accurate simulator oft-used in machine learning research \cite{todorov2012}, which itself doesn't have biomechanical models that compare with the accuracy of OpenSim models. Furthermore, the correspondence between OpenSim and MuJoCo model definitions, in terms of building blocks of the models and their configurations, is high enough to warrant attempts at creating an automatic model converter (see MuJoCo discussion forum for multiple threads on the subject). However, to the best of authors' knowledge, there are no MuJoCo converted models publicly available, nor is there a converter that works with reasonably complex OpenSim models. We present a new converter that is publicly available at \textit{https://github.com/aikkala/O2MConverter} and is able to process complex OpenSim models.


\section{MATERIAL AND METHODS}
\label{sec:methods}

In our experiments we used a complex OpenSim model comprised of a fixed torso and dynamic shoulder and arm \cite{saul2015}. The model has seven degrees of freedom, including shoulder rotation and elevation, elbow flexion, forearm rotation, and wrist flexion and deviation, but we locked wrist flexion and deviation to improve OpenSim simulation stability. The model is actuated by 50 muscles that cover all the remaining five degrees of freedom.



We used OpenSim 4.0 and MuJoCo 2.0 and their Python bindings to run the experiments, all on the same laptop equipped with an Intel i7-8850H processor and 32GB RAM.

\subsection{Model Conversion}
\label{subsec:model_conversion}

Converting OpenSim models to MuJoCo is relatively straightforward: both software use an XML based model definition, and the model parts -- bodies, joints, musculo-tendon units -- are largely equivalent. In fact, building an equivalent skeletal model in MuJoCo is only a matter of disassembling the bodies and joints of an OpenSim model and re-configuring them into a MuJoCo model.

However, there are some anatomical details that are difficult to model exactly in MuJoCo and must be approximated to some extent. For instance, in OpenSim forces acting on joints can be defined in a piecewise linear manner (called \textit{CoordinateLimitForce}), which is cumbersome to model exactly in MuJoCo. OpenSim also offers more flexibility over the anatomical modelling of a musculo-tendon unit (MTU), and particularly problematic are OpenSim's dynamically moving MTU path points whose locations change as a function of a specified joint coordinate value (\textit{MovingPathPoint} and \textit{ConditionalPathPoint}). All these approximations cause inaccuracies in the converted model, which can be mitigated to some extent by optimizing the converted model's parameters.

\subsection{Optimization}
\label{subsec:optimization}

In order to optimize the converted model's parameters, we generated a 100 sets of muscle activations. These activations were modelled as slow frequency (max 1 Hz) sine waves with varying phases, sampled at a frequency of 500 Hz. To increase OpenSim simulation stability, the muscle activation sets had a duration of 1 second, and only a third of muscles were active in a set.

These muscle activations were then used as controls in both OpenSim and MuJoCo forward dynamics simulations -- with a simulation timestep of 2 milliseconds in both simulators -- to generate trajectories. OpenSim's forward dynamics failed to run for 3 sets of muscle activations, and thus we used a subset of 78 trajectories for optimizing the parameters, and a subset of 19 trajectories to estimate joint errors before and after parameter optimization. The parameters that we optimized were: damping and joint limit softness for each non-locked degree of freedom, damping and stiffness for each tendon, and scale (roughly equivalent to strength) of each muscle (160 parameters in total).

The converted model's parameters were optimized with a Python implementation of CMA-ES \cite{hansen2003, hansen2019}, a popular black-box optimization algorithm. We used squared difference of joint positions between OpenSim and MuJoCo forward simulations, summed over all seven degrees of freedom and all training trajectories, as the objective to be minimized. We also augmented the objective with a small cost term to discourage unrealistically high parameter values.


It is important to note that even smallest differences between the models can make their trajectories diverge significantly over time. To alleviate the divergence, we augmented the controls with corrective control signals. These corrective signals were optimized separately for each test set run by minimizing the sum of squared joint position differences, while simultaneously penalising for large corrective signal values using L2 loss. The corrective signals were sampled at 10 Hz and modelled as cubic splines.

\section{RESULTS}
\label{sec:results}

\begin{figure}
\centering
\includegraphics[width=\linewidth]{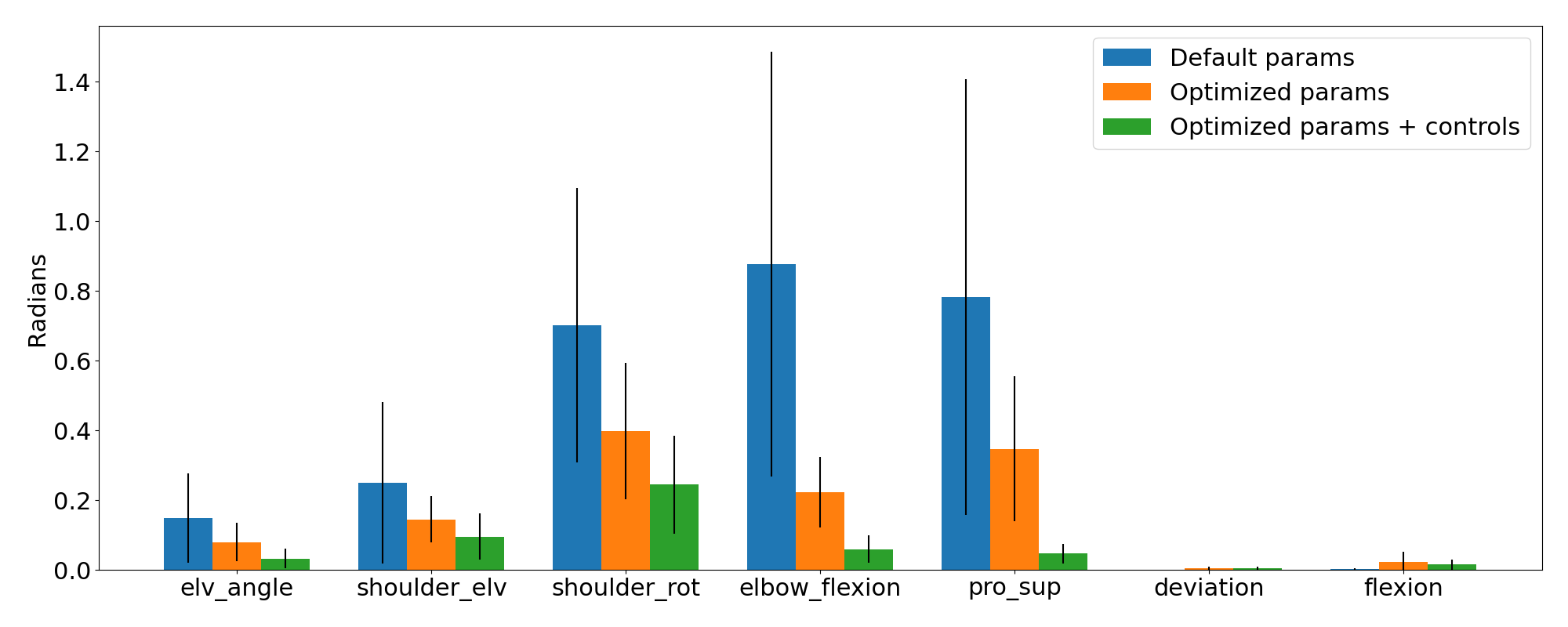}
 \caption{Mean absolute error in radians for each joint over 19 test trajectories.}
\label{fig:joint_errors}
\end{figure}

\begin{figure}
\centering
\includegraphics[width=\linewidth]{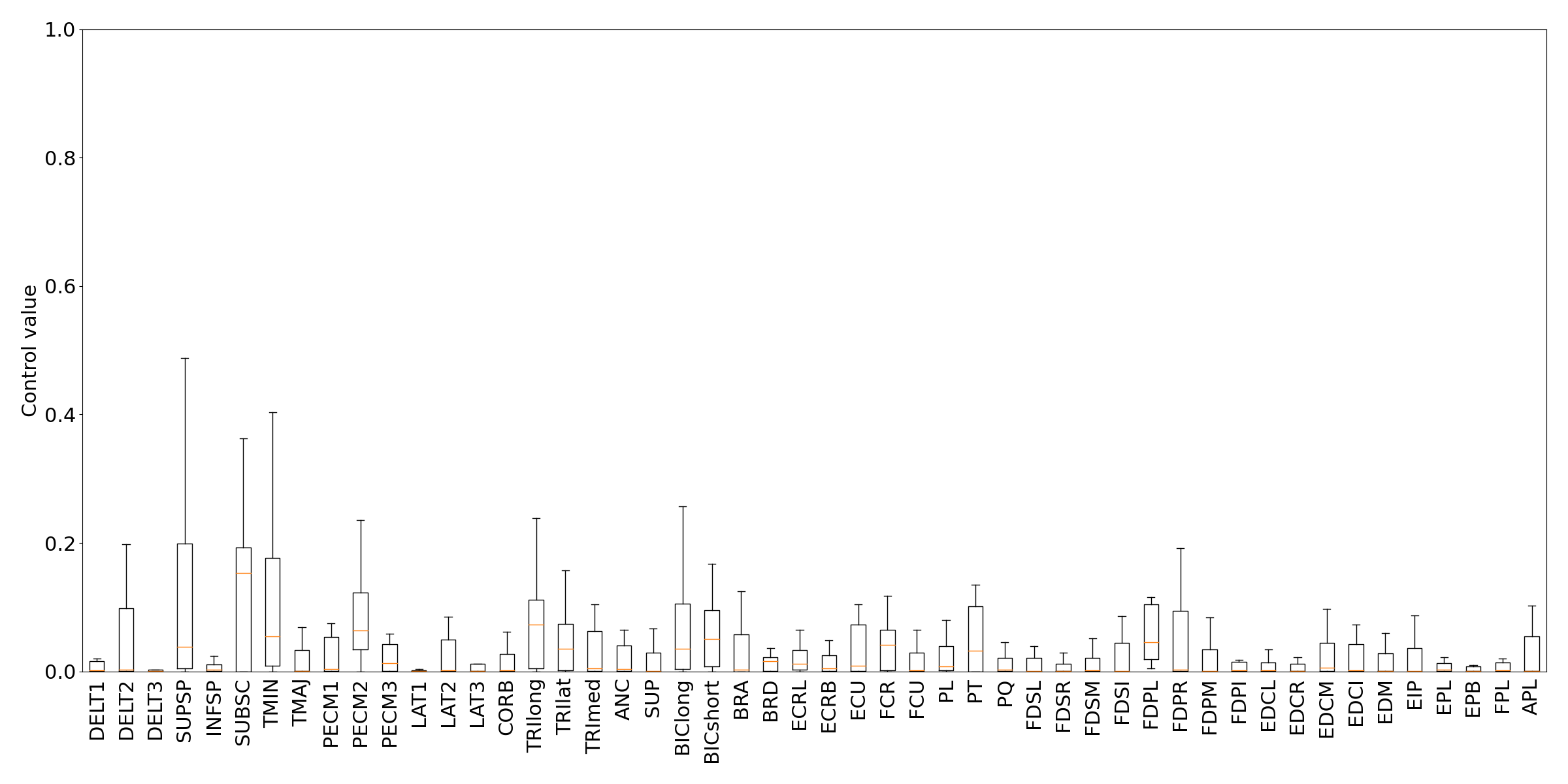}
 \caption{A boxplot depicting mean absolute difference between reference and optimized muscle activations over 19 test trajectories (outliers not visualised). Muscle activations are in range $[0, 1]$, making the maximum difference 1.0.}
\label{fig:control_errors}
\end{figure}

To estimate the accuracy of the converted model, we compare the mean absolute errors for each joint over trajectories in the test set, before and after parameter optimization; see Fig. \ref{fig:joint_errors}. See Fig. \ref{fig:control_errors} for mean absolute difference between reference muscle activations and optimized muscle activations over test set runs.  

A video example of OpenSim reference movement alongside with the converted model's replicated movements, with and without parameter and muscle activation optimization, is available at \textit{https://youtu.be/Nz3R6-1l3lU}.

In addition to estimating accuracy of the converted model, we also compared the efficiency of both simulators by calculating average run time over all 97 forward simulations. One OpenSim simulation took 15.50 seconds on average, while a MuJoCo simulation ran for 0.025 seconds on average, which makes MuJoCo roughly 600 times faster.

\section{Discussion}
\label{sec:discussion}

Fig. \ref{fig:joint_errors} and Fig. \ref{fig:control_errors} indicate that divergence of forward simulation cannot be prevented simply by optimizing the converted model's parameters. Fortunately, OpenSim and MuJoCo simulations can be made almost identical with only minor corrections to the muscle activations. This suggests that MuJoCo should be able to produce reasonably realistic results, e.g., in discovering muscle activation sequences through deep reinforcement learning (for example, see \cite{lee2019}).



Additionally, MuJoCo simulations are substantially faster, and the converted model does behave in a similar fashion to the OpenSim model even with only parameter optimization. This might be enough for some use cases, e.g. in animation and machine learning research. 

Finally, it should be noted, that in addition to accuracy and speed, there's a third performance metric: stability. In order to ensure OpenSim simulations didn't crash we had to lock wrist flexion and deviation, and find suitable starting positions for joints. Even then 3 out of the 100 OpenSim simulations crashed, whereas the MuJoCo model had no problems with wrist flexion and deviation or joint starting positions. This is a major advantage when using such complex biomechanical models for research that requires massive amounts of simulations, such as machine learning research.



\end{document}